\documentclass[aps,prl,twocolumn,superscriptaddress,showpacs]{revtex4}
\usepackage[latin1]{inputenc}
\usepackage{graphicx}
\usepackage{amsmath,amsfonts}

\begin{document}

\title{Scattering of surface plasmons by one-dimensional periodic nanoindented surfaces}

\author{F. López-Tejeira}
\email[Electronic address: ]{flt@unizar.es}
\affiliation{Departamento de Física de la Materia Condensada,
Universidad de Zaragoza, ICMA-CSIC, E-50009 Zaragoza, Spain}

\author{F. J. García-Vidal}
\affiliation{Departamento de Física Teórica de la Materia
Condensada, Universidad Autónoma de Madrid, E-28049 Madrid, Spain}

\author{L. Martín-Moreno}
\affiliation{Departamento de Física de la Materia Condensada,
Universidad de Zaragoza, ICMA-CSIC, E-50009 Zaragoza, Spain}

\begin{abstract}
In this work, the scattering of surface plasmons by a finite
periodic array of one-dimensional grooves is theoretically
analyzed by means of a modal expansion technique. We have found
that the geometrical parameters of the array can be properly tuned
to achieve optimal performance of the structure either as a Bragg
reflector or as a converter of surface plasmons into light. In
this last case, the emitted light is collimated within a few
degrees cone. Importantly, we also show that a small number of
indentations in the array are sufficient to fully achieve its
functional capabilities.
\end{abstract}

\pacs{73.20.Mf, 78.67.-n, 41.20.Jb}

\maketitle

Surface plasmons (SPs) are well known for their capabilities for
concentrating light in sub-wavelength volumes and guiding light
through the surface of a metal \cite{Barnes03}. This has recently
raised the prospect of SP-based photonic circuits, with length
scales much smaller than those currently achieved
\cite{Ditlbacher02,Weeber01,Bozhe01}. In order to reach this goal,
it is mandatory to know the scattering properties of SPs by simple
surface profiles, as pointed out by recent experimental studies
\cite{Ditlbacher02,Devaux03,Weeber04,JGRivas04,Bozhe05,Stepajun05}
From the theoretical side, this is still an open problem, even
after a few seminal works based on the Rayleigh expansion
\cite{SG98,SG99,SG05} and the Green's function dyadic
\cite{Pince94,Sonde}. Both these methods require the evaluation of
sophisticated scattering functions from which physical insight is
not easily inferred.

In this work we present an alternative formalism for calculating
the scattering of SPs by indentations perforated on a thick metal
film. For that purpose, we have extended to real metals the modal
expansion technique previously developed within the perfect
conductor approximation (PCA) \cite{FJ03,JBravo04}, therefore
incorporating surface plasmon polaritons into the model. This
approach enables us to describe the scattering properties of an
arbitrary set of indentations without any restriction over their
position or shape. Besides, our method does not require any
adjustable parameter, inasmuch as the wavelength-dependent
dielectric function of the metal $\varepsilon(\lambda)$ is
previously known. In addition to this, it provides a very compact
representation of the electromagnetic (EM) fields and simple
expressions for the scattering magnitudes, which permits us to
extract underlying physical mechanisms much more easily.

Within our theoretical framework, the way of launching SPs onto a
set of indentations resembles the back-side illumination employed
in some experimental works \cite{Devaux03}. We consider a single
slit flanked by a set of $N$ indentations placed in the output
surface of an infinite metallic film of thickness $h$ (see Fig.1).
Eventually, the distance between the slit and indentations will be
taken to be infinity. In this way, the slit merely plays the role
of a theorist's  SP-launcher. More precisely speaking, SPs are
taken into account by applying surface impedance boundary
conditions (SIBC) \cite{Jack_book} to the metal/dielectric
interface along the film surface.
\begin{figure}[b]
\includegraphics[angle=0,width=8.0cm]{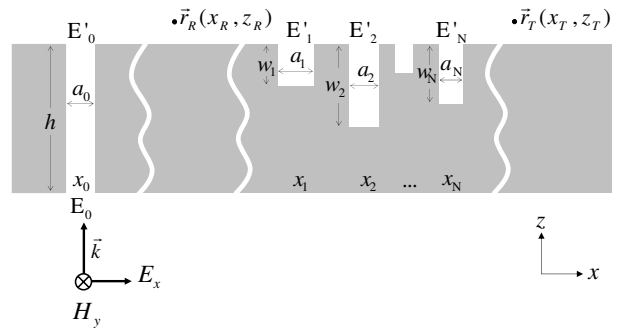}
\caption{Schematic picture of the system under study: single slit
flanked in the output surface by an arbitrary set of $N$
indentations located at a long distance from its right side. A
$p$-polarized EM wave is impinging from the bottom. Parameters
$\lbrace x_{\alpha },a_{\alpha},\mathrm{w}_{\alpha}\rbrace$ define
the geometry of indentations.}\label{virtualslit}
\end{figure}

In this work, we consider the simplest case of 1D subwavelength
indentations (grooves). Additionally, we impose that the external
illumination be uniform along the $y$-axis, so we restrict
ourselves to scattering of SPs impinging onto the grooves at
normal incidence, where only the fundamental mode inside each
groove is relevant. This leads to a set of $N+2$ equations for the
modal amplitudes of the electric field at the input and output
openings of the slit (${E_0,E'_0}$) and the output openings of the
$N$ grooves ($\lbrace E'_{\alpha}\rbrace$):
\begin{equation}
\left.  \begin{array}{rcl}
 (G_{00}-\epsilon_0)E_0 - G_{v 0}E'_0 & = & I_0,\\
  & &  \\
(G_{\alpha\alpha}-\epsilon_{\alpha})E'_{\alpha}+{\displaystyle
\sum_{\beta \neq \alpha}}G_{\alpha \beta}E'_{\beta} - G_{v 0}E_0
\delta_{\alpha 0} & = & 0.
\end{array} \right \rbrace \label{tbeq1}
\end{equation}
Notice that this set of equations is the same as the one obtained
within the PCA \cite{JBravo04}. Therefore, its physical meaning
remains unchanged, although the expressions are slightly different
due to the non-zero impedance $Z_s=\varepsilon(\lambda)^{-1/2}$ at
the metal surface: $I_0$ takes into account the back-side
illumination on the slit (which must be $p$-polarized in order to
excite SPs); the ``self-energy'' $\epsilon_{\alpha}$ is related to
the bouncing back and forth of the EM fields inside indentation
$\alpha$. For a groove,
$\epsilon_{\alpha}=-i(1+\phi_{\alpha})/(1-Z_s-(1+Z_s)\phi_{\alpha})$,
where $\phi_{\alpha}=e^{2ik_0\mathrm{w}_{\alpha}}(1-Z_s)/(1+Z_s)$,
being $k_0 \equiv 2 \pi/ \lambda$ and $\mathrm{w}_{\alpha}$ the
depth of the groove;
$G_{v0}=2i/(e^{ik_0h}(1+Z_s)^2-e^{-ik_0h}(1-Z_s)^2)$ represents
the coupling between the two sides of the slit. Finally,
$G_{\alpha\beta}$ is the coupling between modes, reflecting that
mode $\beta$ emits radiation that can be collected by mode
$\alpha$. More precisely, $G_{\alpha\beta}$ is the projection onto
the wavefields at the openings of indentations $\alpha$ and
$\beta$ of a scalar Green's function
\begin{equation}
G(x,z;x',z') = \frac{i}{\lambda}\int^{+\infty}_{-\infty}dk
\frac{e^{i(k |x-x'|+\sqrt{k_0^2-k^2}|z-z'|)}}{\sqrt{k_0^2-k^2} +
k_0Z_s} .\label{def_g}
\end{equation}
As $Z_s \rightarrow 0$, 
Eq. (\ref{def_g}) transforms into an integral representation of
the 0th-order Hankel function of the first kind \cite{FJ03}, thus
recovering the PCA in the low-energy limit of our approach. Even
within the SIBC approximation, we find that the PCA result is
still valid for $|x-x'|<<\lambda$. However, the presence of $Z_s$
in Eq.(\ref{def_g}) strongly modifies its long-distance behavior.
By means of the saddle-point approximation, we have obtained an
asymptotic expansion of $G$ in the limit where $ |z-z'|,\lambda
\ll |x-x'|$:
\begin{equation}
G_{as}(x,z;x',z') = -(k_0^2 Z_s/k_p)\,e^{i(k_p|x-x'|-k_0 Z_s
|z-z'|)},\label{gasint}
\end{equation}
with $k_p$ satisfying the SP dispersion relation of a flat
metal-dielectric interface within the SIBC,
$\sqrt{k_0^2-{k_p}^2}=-Z_s k_0$. Therefore, the long distance EM
coupling along the surface is due to SPs, even in the presence of
absorption. Comparison with the exact result in the optical regime
shows that the asymptotic limit is already reached for small
distances. For example, in the case of silver at $\lambda= 750$
nm, we find that $G_{as}(\vec{r},\vec{r}\,')$ differs from
$G(\vec{r},\vec{r}\,')$ by less than $10\%$ for $|x-x'| \approx 2
\lambda$. It is this knowledge of long-distance EM coupling being
mediated by plasmons what allows us to use the system in Fig.
\ref{virtualslit} for the analysis of SP scattering.

Let us then examine the term $G_{0 \alpha}E'_{0}$ in Eq.
(\ref{tbeq1}). It can be interpreted as an ``slit illumination''
impinging on the grooves. Thus, the equation governing the EM
fields at the grooves becomes
\begin{equation}
(G_{\alpha\alpha}-\epsilon_{\alpha})E'_{\alpha}+\sum_{\beta \neq
\alpha, 0}G_{\beta
\alpha}E'_{\beta}=\tilde{I}_{\alpha}\label{tbeq3}
\end{equation}
where $\tilde{I}_{\alpha} \equiv -G_{0 \alpha}E'_{0}$ is defined
to resemble the back-side illumination $I_0$. The key point is
that, according to Eq. (\ref{gasint}), $\tilde{I}_{\alpha}$
corresponds to a SP illumination, modulated by a constant factor
that depends on the metal thickness, the intensity of back
illumination and the width of the slit. This factor is not
relevant for the determination of scattering coefficients and the
whole slit can then be treated as a theoretical artifact. 

Once the self-consistent $\lbrace E'_{\alpha}\rbrace$ are
obtained, the calculation of the EM field in all space is
straightforward, and therefore both the emittance $S$ (which is
the fraction of incoming SP energy radiated into vacuum) and its
angular distribution. As the EM coupling between the grooves and a
distant point on the surface is due to SP, we can also obtain the
SP reflection ($r$) and transmission ($t$) amplitudes:
\begin{equation}
r = \sum_{\alpha =1}^N c_{\alpha}e^{ik_p x_{\alpha}}E'_{\alpha},
\,\,\, t = 1+\sum_{\alpha = 1}^N c_{\alpha}e^{-ik_p
x_{\alpha}}E'_{\alpha}.\label{rt}
\end{equation}
where $c_{\alpha}=-i k^2_0
\sqrt{a_{\alpha}}(Z_s/k_p)\mathrm{sinc}\,(k_p a_{\alpha}/2)$ is a
geometrical coefficient associated to each indentation. Notice
that, if absorption is present, $Im [k_p] \ne 0$ and the SP
reflected and transmitted currents depend on the points
$(\vec{r}_{\scriptscriptstyle R},\vec{r}_{\scriptscriptstyle T})$
at which the EM are evaluated, reflecting the absorption loss in
the flat regions of the metal surface. This suggests that the
scattering coefficients should be extracted from the EM fields at
points close to the grooves, although in this case it is difficult
to separate the diffractive contribution from the one due to SPs.
Nevertheless, provided that the grating lengths considered are
shorter than the SP absorption length, absorption can be
neglected. In what follows, we present the results obtained under
such assumption for finite periodic arrays of $N$ rectangular
grooves, patterned on a Ag film.
\begin{figure}[b]
\includegraphics[angle=0,width=8.0cm]{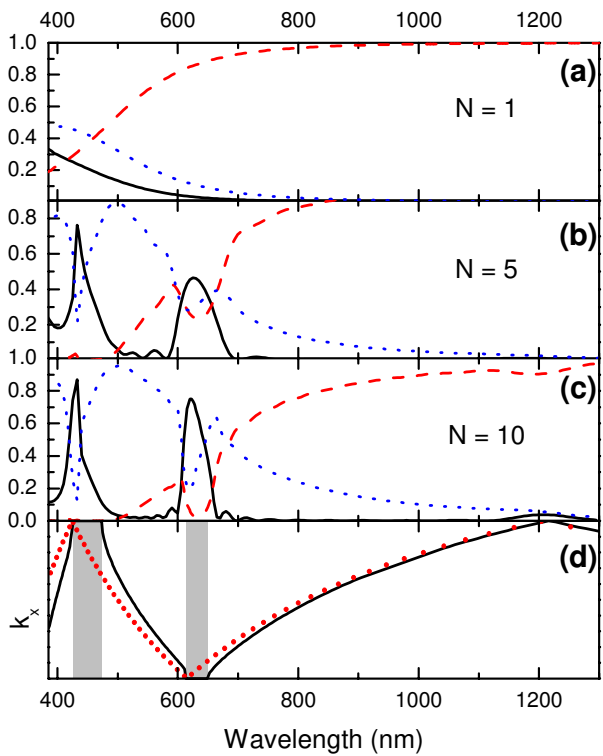}
\caption{(color online). Calculated $R$ (solid), $T$ (dashed) and
$S$ (dotted) curves of SP scattering by a finite periodic groove
array on a Ag film. Here, $a = 100$ nm, $d = 600$ nm and ${\rm
w}=50$ nm. Results in (a), (b) and (c) correspond to structures
with 1, 5 and 10 grooves respectively. Panel (d) shows the  band
structure (solid lines) for the same parameters as in (a) to (c)
and the SP dispersion relation in a flat air/Ag interface (dots).
Gray strips mark the photonic gaps.}\label{RTEbands}
\end{figure}

We consider grooves with width $a=100$ nm and depth ${\rm w}=50$
nm separated by a period $d$ of 600 nm, which are typical
experimental values. Figures \ref{RTEbands}(a)-(c) render the
calculated reflectance $R=|r|^2$, transmittance $T=|t|^2$ and
emittance $S$ spectra, for increasing values of $N$. For a single
groove (top panel), $T$ increases with $\lambda$, while both $R$
and $S$ decrease. This is due to two mechanisms. First, there is a
decrease of the relative size of the groove with respect to
$\lambda$, which manifests in $G_{\alpha \beta}$ scaling as
$(a/\lambda)$. Second, at longer $\lambda$s, the SP wavefield is
more extended in the air region and therefore less  sensitive to
the presence of obstacles at the surface. Panels (b) and (c) show
how the addition of more grooves greatly modifies the optical
response of the system. As $N$ increases, transmission gaps
develop, as well as sharp resonances in both $R$ and $S$.

In order to gain insight into the origin of this behavior, it is
helpful to analyze the EM surface modes of an infinite groove
array. This can be readily done by looking for solutions to Eq.
(\ref{tbeq3}), imposing both $\tilde{I}_{\alpha}=0$ and Bloch's
theorem (i.e. $E'_{\alpha}=E'e^{ik_x \alpha d}$, $k_x$ being the
extended surface state wavevector at the given wavelength). The
band structure (solid line) for surface modes in a periodic
structure with the same geometrical parameters as in (a) to (c) is
presented in Fig.\ref{RTEbands}(d), as well as the dispersion
relation of SPs in a flat air/silver interface (dots). As expected
\cite{Kitson96}, band gaps occur with a low-$\lambda$ edge given
by $k_p d=m\pi$ with $m= 1,2, \cdots$, i.e. by the folding of the
dispersion relation in a flat surface. On the contrary, the
high-$\lambda$ edge depends on the geometry of the grooves, as it
corresponds to a SP standing wave with maxima at the groove
positions. Evidently, spectral regions of low $T$ in the finite
array coincide with gaps in the band structure. Energy
conservation implies a corresponding increase in $R+S$, but it is
not obvious how this increase is divided between these two
channels. However, there is a simple argument for the existence of
reflection maxima. Let us consider the SP wavefields emitted by
two consecutive grooves in the region of reflection. There is an
``optical path" phase difference of $k_p d$ between these waves.
Additionally, there is also a phase difference between emitters
that is equal to $k_x d$ in the case of a infinite system. But, as
previously noted, $k_x d =k_p d = m\pi$ at the low-wavelength gap
edge so the SP wavefields launched by all grooves interfere
constructively (notice that they also interfere constructively in
the transmission region, but not with the incident field). As
$\lambda$ is increased away from this condition, the constructive
interference is progressively lost, $R$ decreases and, for
$\lambda$ within the gap, $S$ increases. If $\lambda$ crosses the
gap edge, the transmission channel is open, and $S$ decreases.
Therefore, $S$ presents peaks at the high-$\lambda$ edges of the
gaps, as can be seen in Fig.\ref{RTEbands}. In our opinion, this
mechanism can explain the heuristic criterion for optimum mirror
efficiency presented in Ref. \cite{Weeber04} and may also be at
the root of the strong asymmetry in the positions of the
reflectance peaks reported in Ref. \cite{Bozhe05}.

\begin{figure}[]
\includegraphics[angle=0,width=8.2cm]{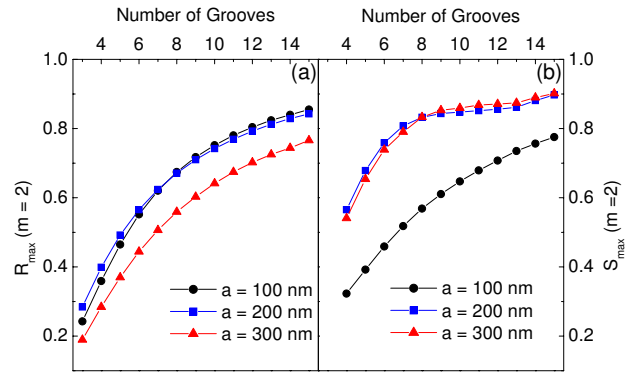}
\caption{(color online). Maximum values of $R$ (a) and $S$ (b) as
a function of the number of grooves in the vicinity of $k_p d =2
\pi$ for the same ${\rm w}$, $d$ as in Fig. \ref{RTEbands} and
increasing values of $a$.}\label{RSmax}
\end{figure}

Figure \ref{RSmax} renders the maximum $R$ and $S$ as a function
of the number of grooves in the vicinity of $k_p d =2 \pi$ for the
same ${\rm w}$,$d$ parameters as in Fig. \ref{RTEbands} and
increasing values of $a$. Notably, a small number of indentations
are sufficient to achieve either a large in-plane reflection or a
high emission out of the plane. This rapid saturation is also
consistent with Ref. \cite{Weeber04}. With respect to the groove
width, it is clear that it mainly influences the out-of-plane
efficiency of every single scatterer, being more relevant for the
$S$ vs $T$ ratio than for mirror efficiency.

\begin{figure}[t]
\includegraphics[angle=0,width=8.0cm]{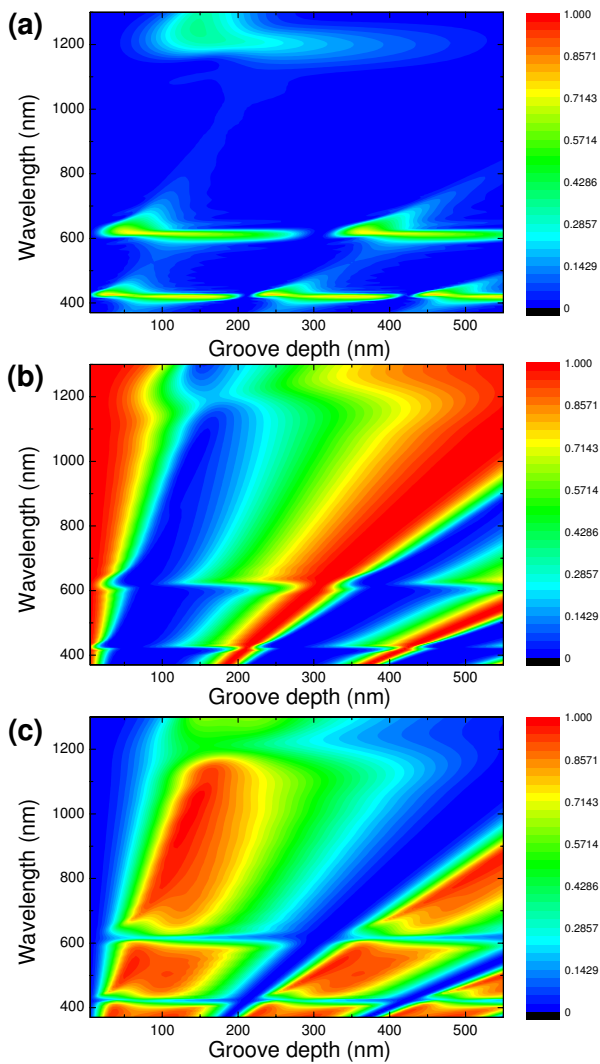}
\caption{(color). Contour plots of $R$ (a), $T$ (b)  and $S$ (c)
versus both groove depth and wavelength for the same ${N,a,d}$ as
in Fig. \ref{RTEbands}(c).}\label{contourp}
\end{figure}

The dependence of the scattering coefficients on groove depth is
depicted in Figure \ref{contourp}, for $N=10$, $a=100$ nm and
$d=600$ nm. As can be seen, the maxima of $R$ vary weakly on $\rm
w$ for most ranges of groove depth. Such a weak dependence may be
relevant for device design, considering that the control of ${\rm
w}$ is often the most difficult point in groove fabrication.
However, Figure \ref{contourp} also shows that for some values of
$\rm w$ the reflection is very small, when a maximum is expected.
Simultaneously, at these groove depths, $S\approx 0$ and $T
\approx 1$. This occurs close to the condition $\lambda=4{\rm
w}/(2n+1)$, when the in-plane electric field at the indentations
is very small, due to destructive interference of the incident
field and the field reflected on the closed end of the groove.

\begin{figure}[b]
\includegraphics[angle=0,width=8.0cm]{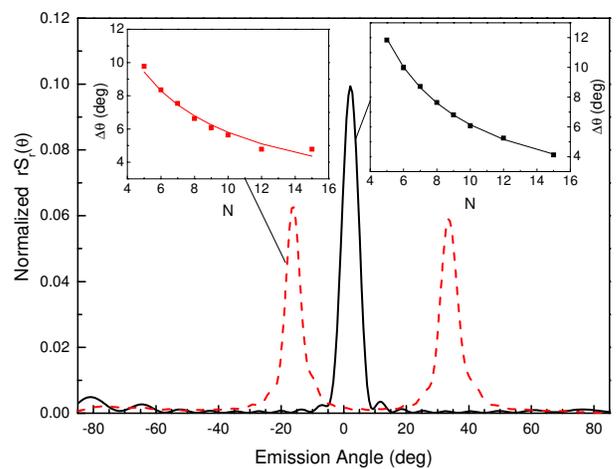}
\caption{(color online). Radial component of the Poynting vector
evaluated in the far field versus angle for an array of grooves
with the same $a,d$,${\rm w}$,$N$ as in Fig. \ref{RTEbands}(c).
The wavelength of the incident radiation corresponds to the $S$
maxima in Fig. \ref{RTEbands}: $\lambda=665$ nm (solid) and
$\lambda=500$ nm (dashed). Top inboxes: FWHM of angular
distribution as a function of $N$ (dots) and $1/N$ fit (solid
lines). }\label{depang} 
\end{figure}

One of the possible applications of finite arrays of indentations
lies in their capability to convert SPs into light. Therefore, it
is worth studying the directionality properties of the emitted
light in the system analyzed throughout this paper. Figure
\ref{depang} shows the far-field angular distribution of radiation
emitted out of plane, evaluated for the two emittance maxima in
Fig. 2c ($\lambda=665$ nm and $\lambda=500$ nm) at the low-energy
edges of the gaps labelled, respectively, with $m=2$ and $m=3$.
The distribution corresponding to the  gap with $m=2$ (solid
curve) is beamed close to the normal. On the contrary, the one
coming from the gap with $m=3$ is beamed at higher angles. Notice
that, at the condition of maximum $R$, ${E_\alpha} \propto
(-1)^{(m \alpha)}$, so the emittance would be normal for even $m$
and close to tangent for odd $m$. Inboxes render the FWHM of
angular distribution peaks for both curves as a function of the
number of grooves. As expected from usual grating theory, it
scales as $1/N$.

To summarize, in this paper we have presented a theoretical
framework that can treat the scattering of SPs by a finite array
of indentations. With this formalism, we have studied the
scattering properties of a sub-wavelength periodic groove array in
the visible and near IR ranges. We have found that, associated to
the low-$\lambda$ edge of the SP band gap, the array behaves as a
mirror (up to 80$\%$ reflectance) whereas at the high-$\lambda$
edge, most of the light carried out by the SP can be converted
into collimated light (up to 90$\%$). We have also shown that this
resonant behavior can be achieved with a small number of
indentations (around 10) and it is quite robust with respect to
variations in the groove depth.

Financial support by the EU (projects FP6-NMP4-CT-2003-505699 and
FP6-2002-IST-507879) and Spanish MEC (contracts MAT2002-01534,
MAT2002-00139 and BFM2003-08532-C02-01) is gratefully
acknowledged.

\begin{thebibliography}{}
\bibitem{Barnes03} See, for example, W. L. Barnes, A. Dereux and T. W. Ebbesen, Nature
(London) {\bf 424}, 824 (2003) and references therein.

\bibitem{Ditlbacher02}H. Ditlbacher, J. R. Krenn, G. Schider, A. Leitner and F. R. Aussenegg , Appl. Phys. Lett.
{\bf81}, 1762 (2002).

\bibitem{Weeber01} J. C. Weeber, J. R. Krenn, A. Dereux, B. Lamprecht, Y. Lacroute and J. P. Goudonnet, Phys. Rev B {\bf 64}, 045411 (2001).

\bibitem{Bozhe01}S. I. Bozhevolnyi, J. Erland, K. Leosson, P. M. W. Skovgaard and J. M. Hvam, Phys. Rev. Lett. {\bf 86},
3008 (2001).

\bibitem{Devaux03} E. Devaux, T. W. Ebbesen, J. C. Weeber and A. Dereux, Appl. Phys. Lett.
{\bf83}, 4936 (2003).

\bibitem{Weeber04} J. C. Weeber, Y. Lacroute, A. Dereux, E Devaux, T. W. Ebbesen, M. U. González and A. L. Baudrion, Phys. Rev B {\bf 70}, 235406 (2004).

\bibitem{JGRivas04} J. Gómez-Rivas, M. Kuttge, P. Haring Bolivar and H. Kurz, Phys. Rev. Lett.
{\bf 93}, 256804 (2004).

\bibitem{Bozhe05} S. I. Bozhevolny, A. Boltasseva, T. S\o ndergaard,T. Nikolajsen and K. Leosson, Opt. Comm. {\bf 250},
328 (2005).

\bibitem{Stepajun05} A. L. Stepanov, J. R. Krenn, H. Ditlbacher, A. Hohenau, A. Drezet, B. Steinberger,
A. Leitner and F. R. Aussenegg, Opt. Lett. {\bf 30}, 1524 (2005).

\bibitem{SG98} J. A. Sánchez-Gil, Appl. Phys. Lett. {\bf 73}, 3509
(1998).

\bibitem{SG99} J. A. Sánchez-Gil and A. A. Maradudin, Phys. Rev.
B {\bf 60}, 8359 (1999).

\bibitem{SG05} J. A. Sánchez-Gil and A. A. Maradudin, Appl. Phys. Lett. {\bf 86}, 251106
(2005).

\bibitem{Pince94} F. Pincemin, A. A. Maradudin, A. D. Boardman and J. J. Greffet, Phys. Rev. B {\bf 50},
15261 (1994).

\bibitem{Sonde} T. S\o ndergaard and S. I. Bozhevolnyi, Phys.
Rev. B {\bf 67} 165405 (2003); {\bf 69} 045422 (2004); {\bf 71}
125429 (2005).

\bibitem{FJ03} F. J. García-Vidal, H. J. Lezec, T. W. Ebbesen and L. Martín-Moreno, Phys. Rev. Lett.
{\bf 90}, 213901 (2003).

\bibitem{JBravo04} J. Bravo-Abad, F. J. García-Vidal and L. Martín-Moreno, Phys. Rev. Lett. {\bf 93}, 227401
(2004).

\bibitem{Jack_book} J.D. Jackson, {\it Classical Electrodynamics}, 2nd ed., Wiley, New York (1975).

\bibitem{note_bozhe} Also found in \cite{Sonde} beyond the range of validity of SIBC.

\bibitem{Kitson96} S. C. Kitson, W. L. Barnes and J. R. Sambles, Phys. Rev. Lett.
{\bf 77}, 2670 (1996).



\end{thebibliography}
{}

\end{document}